\documentclass[aps,pra,preprintnumbers,amsmath,amssymb,showpacs]{revtex4}
\usepackage{graphicx}
\usepackage{dcolumn}
\usepackage{bm}
\usepackage{hyperref}
\usepackage[latin1]{inputenc}
\usepackage{pstricks}           

\newcommand{\eps}{{\bf \varepsilon}}

\begin{document}

\title{Attosecond streaking of photoelectrons emitted from metal surfaces}

\author{Boyan Obreshkov}

\affiliation{Institute for Nuclear Research and Nuclear Energy,
Bulgarian Academy of Sciences, Tsarigradsko chausse\'{e} 72, Sofia
1784, Bulgaria}{1}

\begin{abstract}
We numerically investigate attosecond streaking time delays in the
photoemission of valence and 2p core electrons of aluminum
surface. We find that electron emission from the core level band
is delayed by $\Delta \tau =100$ attoseconds relative to the
release of electrons from the valence band. We show that this
relative time offset in electron emission is caused by the
screening of the streaking laser field by conduction electrons.
\end{abstract}

\maketitle

\section{Introduction}
The progress in development  of ultrashort laser pulses has
allowed time-resolved study of photoemission on atomic time scale.
In attosecond time-resolved measurements of photoelectron emission
from metal surfaces, attosecond ($10^{-18} s$) extreme ultraviolet
(XUV) light pulse is used to ionize electrons from either bound
core levels or delocalized conduction-band states. The released
electrons move toward the surface, where they are exposed to an
intense few-cycle infrared (IR) laser pulse at the moment of their
escape from the surface. The electric field of the IR pulse
modulates the kinetic energy of the electrons depending on their
release time from the crystal. Although valence and core electrons
are ionized at the same time, it was experimentally found that
they emerge from the surface at different times, therefore their
measured kinetic energies showed a small relative time shift
$\Delta \tau$. In this experiment \cite{exp_atto_tung}, the
emission of 4f core electrons of tungsten was found to be delayed
relative to the valence band by $110 \pm 70$ attoseconds.

This small time difference has been explained with several
different theoretical models. Classical transport simulations
\cite{th_atto_1} interpreted this time delay as the difference in
the average travel times of electrons emerging from delocalized
valence-band and tightly bound core states to the surface of the
crystal. In contrast, quantum mechanical calculations suggest that
the observed time delay is due to the different degree of spatial
localization of initial state wave functions
\cite{th_atto_2,th_atto_3}.

Time resolved photoionization of (free-electron-like) Mg(0001)
surfaces \cite{exp_atto_Mg} found no relative time delay in the
streaked electron emission from localized 2p core and valence band
states of Mg. The authors suggested that transport-related
contributions to attosecond time delays can be explained in terms
of quotient of group velocities and mean free paths of
photoelectrons during their propagation inside the bulk,
regardless of the different degree of localization of initial
state wave-functions.

By solving the time-dependent Schrodinger equation with realistic
1D model potentials describing the electron-metal interactions,
Ref. \cite{th_atto_Mg1} explained this vanishing delay as a result
of dominant electron emission from the bulk valence bands of Mg
via resonant interband transitions. The sensitivity of relative
time delays to the time-dependent response of the substrate
electronic structure was further investigated and elaborated in
Ref.\cite{th_atto_Mg2}, in particular the experimental observation
in  Ref.\cite{exp_atto_Mg} was reproduced provided the
energy-dependence of the photoelectron mean free path and the
screening of the IR field at the surface are properly described.

Common to these model calculations is the description of the
electronic structure by 1D model potentials, thereby assuming
uniform motion of the photoelectrons parallel to the surface.
Furthermore the screening of IR field by conduction electrons was
described phenomenologically in terms of classical skin depth. The
purpose of this paper is not to refine the theoretical
description, but to investigate to what extent screening of the IR
field by conduction electrons affects relative time delays in
photoemission. The paper is organized as follows, In Sec.2 we
present our theoretical model, Sec.3 discusses numerical results
on streaking time delays in the photoemission from free-electron
jellium Al surface and Sec.4 contains our main conclusions.

\section{Theoretical model}

In one electron approximation, the time-dependent Hamiltonian for
a free-electron like metal surface interacting with
electromagnetic XUV and IR fields can be written as
\begin{equation}
H(t)=H_0+H_{{\rm int}}(t)
\end{equation}
and $H_0=T+V_{{\rm eff}}$ is the field-free Hamiltonian given in
terms of an effective one-electron potential. In velocity gauge
the interaction Hamiltonian is
\begin{equation}
H_{{\rm int}}(t)= V_X(t) +V_{{\rm IR}}(t) = A_X(t-\tau)p_z +
\frac{1}{2} (A_z(z,t) p_z +p_z A_z(z,t) +A^2_z(z,t))
\end{equation}
where $p_z=-i \partial_z$ and $\tau$ is the time delay between the
IR and XUV laser pulses. The temporal profile of the XUV pulse is
modelled by a Gaussian function
\begin{equation}
A_X(t) \sim e^{-\ln(4) (t/\tau_X)^2} e^{-i \omega_X t},
\end{equation}
with a pulse duration $\tau_X=432$ as, central frequency
$\omega_X=118$ eV. The IR streaking field is approximated by its
classical form
\begin{equation}
A_z(z,t)=\theta(z) A(t) \label{acl},
\end{equation}
where $\theta$ is the Heaviside step function, Gaussian envelope
of the IR laser pulse is assumed
\begin{equation}
A(t)=A_0 \sin(\omega_L t) e^{-\ln(4) (t/\tau_L)^2}
\end{equation}
with pulse duration $\tau_L=200$ a.u. and driving frequency
$\omega_L=1.5$ eV, the amplitude $A_0$ is chosen such that the
peak intensity of the IR pulse is $I \approx 10^{12} {\rm
W/cm^2}$. The approximation of Eq.(\ref{acl}) relies on the
assumption of abrupt change of screening properties at the
metal-vacuum interface, which is reasonable since the laser
frequency $\omega_L$ is well below the bulk plasmon excitation
threshold. Though it is known that the detailed spatial behavior
of the electromagnetic potential within a few angstrom of the
metal surface is relevant \cite{RPA_scr}, our independent
investigation based on the random phase approximation shows that
the use of non-analytic vector potential is well justified for a
high density substrate such as Al.

Treating the interaction with the XUV pulse as a perturbation, the
time evolution operator is
\begin{equation}
U(t,t_0)=U_1(t,t_0) -i \int_{t_0}^t dt' U_1(t,t') V_X (t')
U_1(t',t_0),
\end{equation}
where $U_1$ describes the propagation of the electron in the
screened IR field, which is formally given by the chronological
Dyson's exponent
\begin{equation}
U_1(t,t_0)=T\exp\left(-i \int_{t_0}^t H_1(t) dt \right),
\end{equation}
where $H_1(t)=H_0+V_{{\rm IR}}(t)$. The photo-ionization amplitude
from an initial target state $|i \rangle$ to a final continuum
state $|f\rangle$ with kinetic energy $E_f$
\begin{equation}
T_{fi} = \langle f | U_1(+\infty,-\infty) | i \rangle -i \int dt
\langle f| U_1(+\infty,t) V_X(t) U_1(t,-\infty) | i \rangle
\end{equation}
includes a background contribution describing multi-photon
ionization process driven by the IR field (first term), and a
resonant contribution associated with the emission of electrons
into the laser-dressed continuum of final states (second term). We
neglect the first term and approximate the transition amplitude by
\begin{equation}
T_{fi}(\tau) \approx -i \int_{-\infty}^{\infty} dt A_X(t)
p_{fi}(t) ,
\end{equation}
where $p_{fi}(t)=\langle \phi_f(t) | p_z | \phi_i(t) \rangle$ is
the off-diagonal matrix element of the momentum, the streaked
photoemission probability is given by the incoherent sum over
initially occupied substrate states
\begin{equation}
\label{trace}  P_f(\tau)= P(E_f,\tau)= \sum_i n_i
|T_{fi}(\tau)|^2.
\end{equation}
After spectrally averaging the streaking traces in
Eq.(\ref{trace}), we obtain time delays in photoemission from the
relative displacement of the kinetic energy centroids $\langle
E_f(\tau) \rangle = \int dE_f E_f P(E_f,\tau) / \int dE_f
P(E_f,\tau)$. To specify the final state wave-function, we
numerically solve the Schr\"{o}dinger equation in momentum space
by a wave-packet propagation method (cf. Appendix)
\begin{equation}
i \partial_t \phi_f(p,t) = \frac{1}{2} p^2 \phi_f(p,t) +
\frac{i}{2} A(t) \int_C \frac{dp'}{2 \pi} \frac{p+p'+A(t)}{p-p'+i
\eta} \phi_f(p',t)
\end{equation}
where $\eta > 0$ is a positive infinitesimal and the contour $C$
is along the real axis. We model the initial state wave-functions
of weakly bound valence electrons by jellium wave-functions with
binding energies $\eps_k=(k^2-k^2_F)/2$, where $k_F$ is the Fermi
momentum
\begin{equation}
\phi_k(p,t)= e^{-i \eps_k t} \left( \frac{1}{p-k+i0}+ \frac{e^{2 i
\eta_k}}{p+k+i0} - 2 \frac{e^{i \eta_k} \cos \eta_k}{p-i \kappa}
\right),
\end{equation}
$\eta_k=-\arctan(\kappa/k)$ is a reflection phase,
$\kappa=\sqrt{2(W-\eps_k)}
>0$ is the
surface barrier penetration coefficient and $W=4$ eV is the Al
workfunction. We assume simple tight-binding approximation for the
description of the initial 2p core-level wavefunctions
\begin{equation}
\phi_{\theta}(p,t)= e^{-i \eps_{2p} t} \Delta_{\theta}(p)
u_{2p}(p),
\end{equation}
in terms of planar-averaged structure factor
\begin{equation}
\Delta_{\theta}(p)= \sum_{l=1}^{N_l} \sin(l \theta) e^{i p z_l},
\end{equation}
where $z_l=-(l-1/2)a_s$ labels the positions of the bulk atomic
layers relative to the jellium edge $(z=0)$ of the surface and
$a_s=3.8$ is the interlayer spacing for the Al(100) surface. The
phase parameter $\theta= k a_s$ changes in the first Brillouin
zone $ 0 \le \theta \le \pi $. The atomic wavefunction $u_{2p}(p)$
is an eigenfunction of Yukawa-type screened Thomas-Fermi potential
with binding energy $\eps_{2p}=72$ eV relative to the Fermi level
$(\eps_F=0)$.

\section{Numerical results and discussion}

\begin{figure}[hb]
\centering
\includegraphics[scale=0.8]{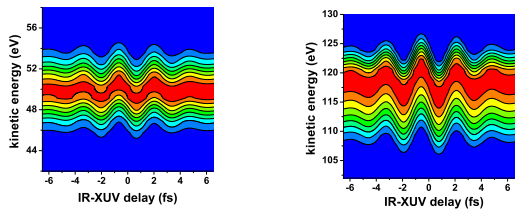}
\caption[]{Streaking spectrograms $P(E_f,\tau)$ of the valence
band (a) and 2p core-level band (b) of an Al(100) surface as a
function of the IR-XUV pulse delay $\tau$ and kinetic energy of
released electrons into the screened IR field. Negative time delay
$(\tau < 0)$ corresponds to XUV preceding the IR pulse.}
\label{fig:trace}
\end{figure}

\begin{figure}[h]
\centering
\includegraphics[scale=0.8]{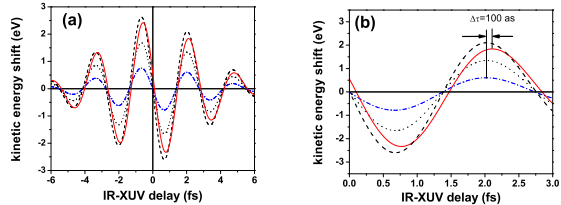}
\caption{Kinetic energy shifts for streaked photoemission from
Al(100) surface. In Fig.(a) and Fig.(b) black dashed line gives
result for the valence electron emission into the unscreened
streaking field, solid red line - release of valence electrons
into the screened IR field, black dotted line -- emission of Al-2p
core electrons into the unscreened potential and blue
dashed-dotted line - release of Al-2p core electrons into the
screened streaking field.} \label{fig:shift}
\end{figure}

The streaked photoemission spectra of valence and 2p core level
bands of aluminum are shown in Fig.\ref{fig:trace}. The
spectrograms display the modulation of the photoelectron kinetic
energy as a function of the IR-XUV time delay $\tau$. The
centroids of the streaked emission spectra follow the time
variation of the IR field, as shown in Fig.\ref{fig:shift}.
Noticeably photoelectrons released from the 2p core level
experience much weaker kinetic energy shifts by interacting with
the IR field. More importantly the relative displacement of the
streaking curves corresponds to a time delay of $\Delta \tau
\approx 108$ as in the emission of 2p core electrons relative to
the valence band. This time offset is in very good quantitative
agreement with the experimental data in Ref.\cite{exp_atto_tung}.
For comparison streaking curves for photoemission into a spatially
uniform IR vector potential are also shown. In this case kinetic
energy shifts of released valence and core electrons are strictly
synchronized in time, i.e. there is no relative time delay in
their emission. Since the time-dependent response of valence
electrons to the screened vector potential causes their advanced
emission relative to their emission in the unscreened vector
potential, this comparison suggests that the time shift of $100$
as is exclusively due to propagation effects of released electrons
into the screened IR field, moreover in this case the initial
state wave-functions $\phi_i(p)$ involved in the transition are
unchanged.

\begin{figure}[h]
\centering
\includegraphics[scale=0.8]{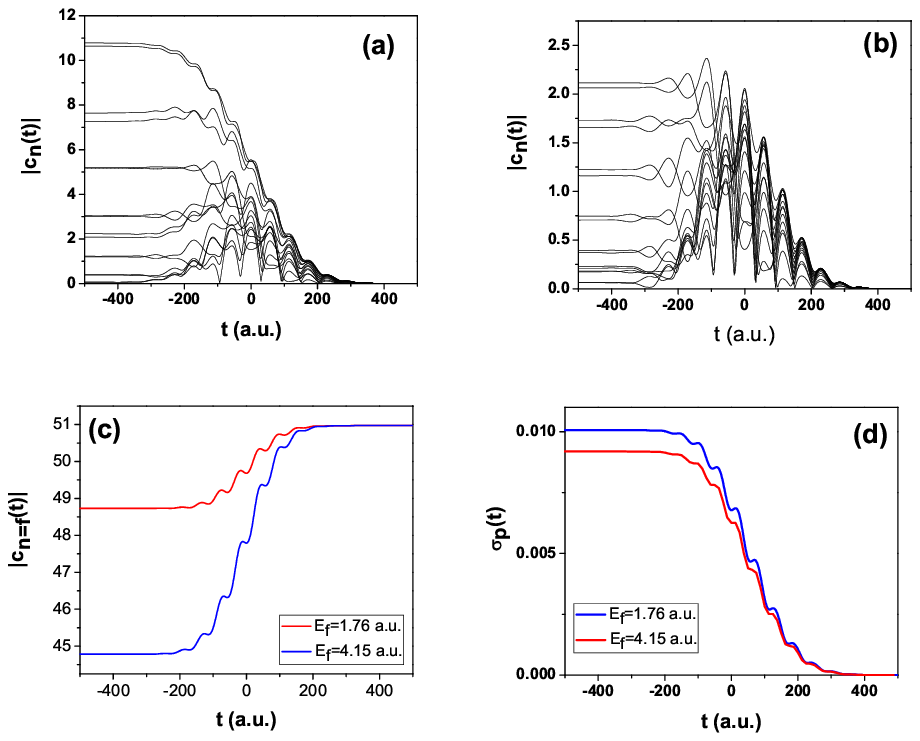}
\caption[]{Time evolution of the Fourier coefficients in the
expansion of the wave-function over plane waves. In Fig. (a) the
kinetic energy of the released electron is $E_f=4.15$ a.u. and it
is $E_f=1.76$ a.u. in Fig. (b).  For these two specific final
kinetic energies Fig.(c) gives the time evolution of the central
momentum components of the two wave packets. Fig.(d) gives the
temporal variation of the mean squared fluctuation of the
photoelectron momentum $\sigma_p=\sqrt{ \langle p^2 \rangle -
\langle p \rangle^2}$.} \label{fig:coeff}
\end{figure}

To anlayze these results further, in Fig.\ref{fig:coeff}(a-d) we
plot the time evolution of the Fourier coefficients $c(E_f,p_n,t)$
in the expansion of the photoelectron wave-function over
plane-waves (cf. Appendix). Fig.\ref{fig:coeff}(a-b) compares the
time development of the IR-field induced admixtures to the central
momentum $p_f=2.8$, Fig.(a), and $p_f=1.8$ in Fig.(b). Propagating
backwards toward the remote past, the coupling to the IR field
gradually builds-up the photoelectron wave-packets by adding up
coherently contributions of nearby momentum eigenstates. The
wave-packets are fully formed for large negative times $t \le
-200$ (remote past), when the Fourier amplitudes change
harmonically over time $c(E_f,p,t)= c(E_f,p) e^{-i p^2/2 t}$. As
this comparison suggests the build-up of the wave-packet is less
efficient for electrons released into lower kinetic energy states.
Fig.\ref{fig:coeff}(c) shows the time evolution of the central
component $c(E_f,p_f,t)$ of the wave-packet, the mean-squared
fluctuation of the carrier momentum $\sigma_p = \sqrt{\langle p^2
\rangle - \langle p \rangle^2}$ is also shown in
Fig.\ref{fig:coeff}(d). The momentum spread over nearby states
leads to uncertainty $\sigma_p \le 0.01$ a.u., i.e. $\sigma_p \ll
p_f$, so that the Fourier coefficients are only appreciable in a
narrow range near the mean momentum $p_f$ during the whole time
evolution.

\begin{figure}[h]
\begin{center}
\includegraphics
{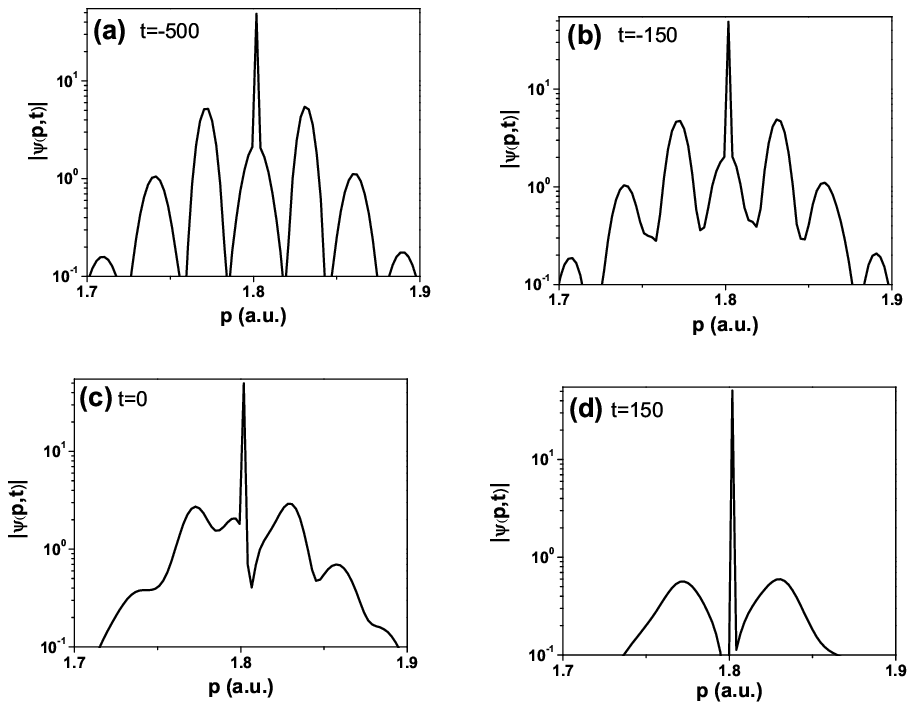}
\caption{Time evolution of the modulus of a laser-dressed
wave-function $|\phi_f(p,t)|$ of a photoelectron released with
kinetic energy $E_f=1.8$ a.u.} \label{fig:wfp}
\end{center}
\end{figure}

The time evolution of the modulus of the momentum-space
wave-function for the lower kinetic energy state is shown in
Fig.\ref{fig:wfp}(a-d). The wave packet is fully formed in the
distant past $(t=-500)$, when density of states is peaked at $p_f$
and is structured by narrow sidebands equidistantly spaced by
$\delta p=0.03$ a.u. above and bellow $p_f$. The number of
sidebands depends sensitively on the number of half-cycles of the
driving pulse. When the time increases to $t=-150$,
Fig.\ref{fig:wfp}(b) the sidebands are distorted and slightly
overlap. At the peak intensity of the IR electric field ($t=0)$,
Fig.\ref{fig:wfp}(c), the sideband structure is blurred showing
relevance of transient effects in the electron dynamics. The
change in the density of states causes reduction of the
wave-packet for large positive times $t=150$,
Fig.\ref{fig:wfp}(d), which is because electrons emerge in the
low-intensity tail of the IR pulse and can undergo only weak
change of their momenta. In the far future $t \rightarrow \infty$,
the components of the wave-packet are projected out describing the
final free-electron state with momentum $p_f$ at the detector.

\begin{figure}[h]
\begin{center}
\includegraphics
{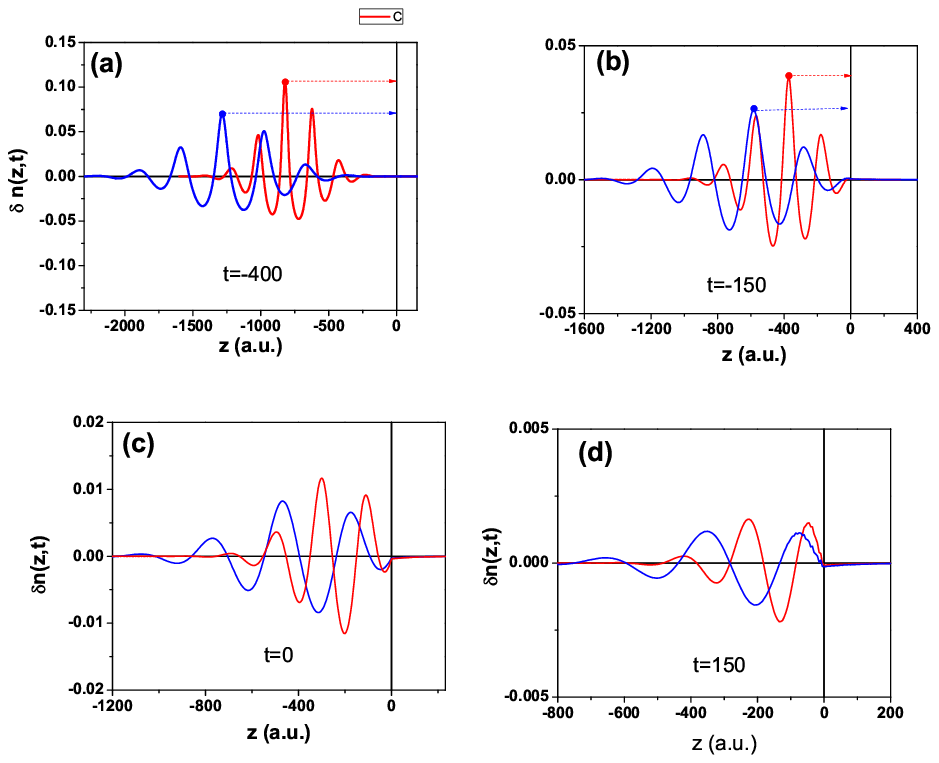}
\caption{Time evolution of the streaked photoelectron density
$\delta n(z,t)=|\psi(z,t)|^2-1$ relative to the uniform
free-electron density, here $\psi(z,t)$ is the (inverse) Fourier
transformation of $\phi(p,t)$ and $z$ is the coordinate normal to
the surface. The metal occupies the half-space $z<0$ and $z>0$
designates the radiation filled vacuum half-space. The blue curves
correspond to electrons released from the valence band and red
curves to photoelectrons released from the 2p core-localized
states.} \label{fig:wpp}
\end{center}
\end{figure}

Fourier transforming these results to coordinate representation,
snapshots of the streaked photoelectron density relative to the
free electron density $\delta n(z,t)= |\psi(z,t)|^2-1$ are shown
in Fig.\ref{fig:wpp}(a-d). If electrons were promoted into the
laser dressed continuum during the remote past $t =-500$,
Fig.\ref{fig:wpp}(a), the IR pulse has ceased and thus
wave-packets are fully formed. Their charge density distributions
display characteristic longwavelength spatial modulation inside
the bulk (with a wavelength of about 20 nanometers) . The crests
of the wave-packets move freely towards the surface following the
classical path $z = -z_{in} + v_{in} t$ with mean velocity
$v_{in}$, but lag behind by a distance $z_{in} >0$ relative to the
free-electron motion. The speeds of the electrons at the instant
of ionization are $v_{in}=2.801$ and $v_{in}=1.799$ a.u.,
respectively. The space shift $\Delta z$ between the crests
corresponds to a relative time delay $t_0=-z_{in}/v_{in}$, which
is $t_0 \approx -1400$ as for the higher energy state, while for
the lower energy one $t_0 \approx -1380$ as. If the uniform motion
of the crests is extrapolated into the surface region $z=0$, it
would suggest that electrons released into lower kinetic energy
states arrive at the surface slightly earlier by $\Delta \tau =20$
as. However this classical estimation is inconsistent with the
time shift of $100$ as we obtain from the displacement of the
streaking curves in Fig.\ref{fig:shift}. The naive time definition
based on the movement of the crest is too crude to account for the
spreading and re-shaping of the wave-packets as electrons move
towards the surface. The final speeds are $v_f=2.830$ and
$v_f=1.801$, respectively, showing that the higher energy state is
accelerated by the IR field $\Delta v= v_f-v_{in}=0.03$ a.u.,
while electrons released into lower kinetic energy states
experience negligible velocity shifts.

If electrons were ionized at a later time $t=-150$,
Fig.\ref{fig:wpp}(b), the crests of the wave-packets have moved
closer to the surface, the spatial profile of the bulk charge
density modulation changes due to reshaping of the wave-packets.
If electrons are ionized at the peak intensity of the IR pulse,
Fig.\ref{fig:wpp}(c), the free-electron motion is strongly
perturbed by the IR field causing sharp truncation of the charge
densities at the jellium edge. Such rather abrupt variation of the
density near the surface is because photoelectrons increment their
momenta $p \rightarrow p+A(t)$ after passing into the
radiation-filled half-space $(z>0)$. This can be formally
expressed by the jump of the derivative of the wave-function at
the jellium edge $z=0$, i.e.
\begin{equation}
\frac{\partial \psi}{\partial z}(z=0^+,t)-\frac{\partial
\psi}{\partial z}(z=0^-,t) =-i A(t) \psi(z=0,t).
\end{equation}
Though the ordinary derivative has a discontinuity at $z=0$, the
covariant derivative $[\partial_z+iA(z,t)]\psi(z,t)$ remains
continuous across the metal-vacuum interface. The electron density
converges to a spatially uniform distribution for large positive
times $t=150$ (far future), Fig.\ref{fig:wpp}(d), such that
wave-packets are dissolved and reduce to a single plane wave state
$|\psi(t \rightarrow \infty) \rangle \rightarrow |p_f \rangle$
describing the emitted electron.

\begin{figure}[h]
\begin{center}
\includegraphics
{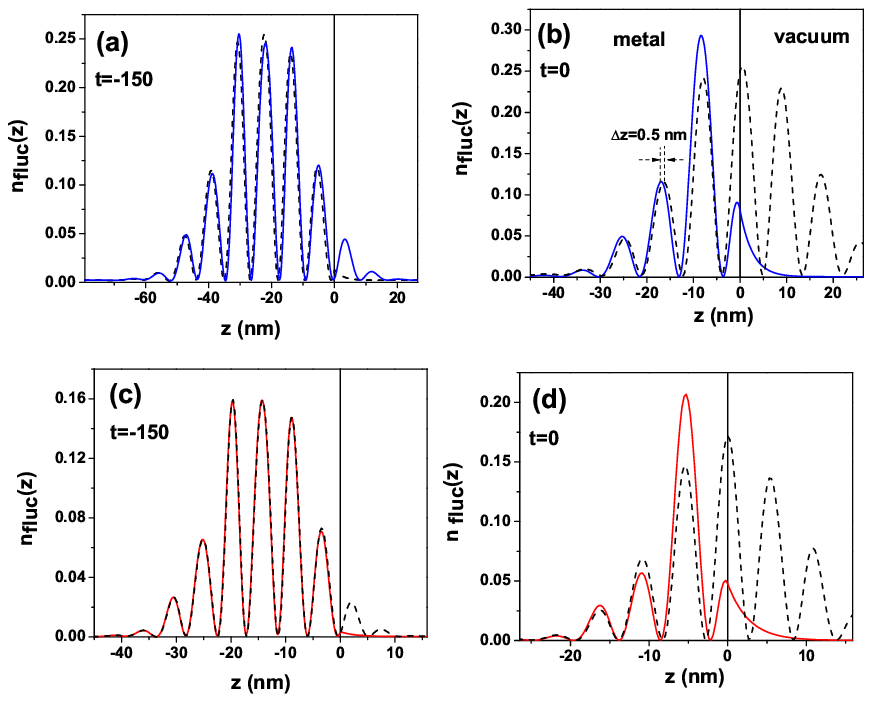}
\caption{Time evolution of the sideband structure of streaked
photoelectron wave-packet. The dashed and solid lines in (a-d)
give the field-free ($A=0$) and field-dependent time evolutions,
respectively. Fig.(a) and (b) give density distributions for
photoelectrons released with kinetic energy $E_f=4.15$ at times of
ionization $t=-150$ and $t=0$, respectively. Fig.(c) and (d) show
the distributions of photoelectrons released with kinetic energy
$E_f=1.76$ at times of ionization $t=-150$ and $t=0$,
respectively. The distance to the surface is given in nanometers,
the metal occupies the half-space with $z<0$ and $z
>0$ is the radiation-filled half-space). } \label{fig:re}
\end{center}
\end{figure}

To further analyze temporal changes of photoelectron
distributions, it is more instructive to decompose the
photoelectron wave-function according to
\begin{equation}
\psi(z,t)=a(t) e^{i p_f z} + \sum_{p_n \ne p_f} \sqrt{w_n}
c(p_n,t) e^{i p_n z} = \psi_f(z,t) + \psi_{fluc}(z,t)
\end{equation}
where the first term is a renormalized plane-wave of the
photoelectron with mean momentum $p_f$ and the momentum spread
over nearby states is represented by the fluctuating part of the
wave-function $\psi_{fluc}$. In Fig.\ref{fig:re}(a-f) we plot the
time-evolution of the fluctuating part of the charge density
$n_{fluc}(z,t)=|\psi_{fluc}(z,t)|^2$ for electrons released into
the screened IR field,  for comparison the field-free time
evolution is also shown. The field-free wave-packet is given by
\begin{equation}
\psi_0(z,t)= a_0(t\rightarrow -\infty) e^{i p_f z} e^{-i p_f^2/2
t} + \sum_{p_n \ne p_f} \sqrt{w_n} c(p_n,t \rightarrow -\infty)
e^{i p_n z} e^{-i p_n^2 t/2}.
\end{equation}
Both $\psi$ and $\psi_0$ satisfy the free Schr\"{o}dinger equation
inside the metal $z<0$. As Fig.(\ref{fig:re})(a) demonstrates, the
wave-packet of the sideband momentum eigenstates is spatially
localized and is represented by a normalized wave-function. Inside
the metal, the free- and streaked- electron distributions display
interference fringes that are slightly displaced on a
sub-nanometer length scale. Both wave-packets are travelling in
the direction of the positive $z$-axis (the vacuum half-space) at
a constant speed. However unlike the freely moving wave-packet,
the front-end of the streaked wave-packet is truncated inside
vacuum, i.e. $\psi_{fluc}$ has no propagating component in the
radiation-filled half-space but only slowly decaying evanescent
one. At the peak intensity of the IR pulse $(t=0)$, cf.
Fig.(\ref{fig:re})(b), the wave-packet is strongly distorted, the
front-end part is truncated inside vacuum (relative to the free
motion), the probability density remains localized at the
metal-vacuum interface. Inside metal $(z<0)$, where the electric
field of the IR pulse is fully screened, the back-end of the
wave-packet is spatially shifted by $\Delta z = 0.5$ nm relative
to the free motion. This space shift corresponds to relative time
delay $\Delta \tau= \Delta z/ v_f \approx 100$ as in the release
of electrons into the streaking field. The direction of the shift
shows that photoelectrons move at a higher speed in the radiation
filled half-space and are thus emitted in advance (relative to the
free-electron motion). As this comparison demonstrates, the time
shift we get in this way is in agreement with the timing
information we get from the relative shift of the streaking traces
in Fig.\ref{fig:trace}. This confirms our observation that change
in propagation of photoelectrons upon screening of the
electromagnetic field by conduction electrons gives dominant
contribution to the relative time delay in the photoemission.

Similar results are obtained for electrons released from the
2p-core level states of Al, Fig.(\ref{fig:re})(c-d).  The
difference is that the relative phase shift inside the metal
caused by the interaction with the IR field is vanishingly small,
resulting in no noticeable relative time delay in the emission of
core electrons (relative to the free motion).

\section{Conclusion}

Based on wave-packet propagation study, we have investigated
attosecond time delays in the photoemission from Al surfaces. We
find that 2p core electrons are emitted with delay relative to the
valence band by $100$ attoseconds. We find that screening of the
IR field is essential determinant for the calculation of the
relative time delays in the emission. This relative time offset in
the emission results from electron propagation effects (spreading
and re-shaping of the wave-packets) in the screened electric field
of the IR pulse. Though our numerical result does not include
corrections to the streaking time shifts due to electron-electron
and electron-ion collisions inside the bulk, we find the result is
in very good quantitative agreement with the experiment. More
detailed investigation of these effects is a subject to our
follow-up paper. We also hope this work may become helpful in
interpretation of experimental data on attosecond time-resolved
photoemission from metal surfaces.

\section*{Appendix}

The Schr\"{o}dinger equation $i \partial_t \phi(t)=H(t) \phi(t)$
is discretized in momentum space using Gauss-Legendre mesh points
$p_n$ with weights $w_n$ (\cite{DVR})
\begin{equation}
i \partial_t c_n(t) = \sum_m H_{nm}(t) c_m (t) ,
\end{equation}
and the definitions
\begin{equation}
c_n(t)=\sqrt{w_n} \phi_f(p_n,t),
\end{equation}
where
\begin{equation}
H_{nm} (t) = \frac{1}{2 \pi} \sqrt{w_n w_m} H(p_n,p_m,t),
\end{equation}
is the matrix representation of the Hamiltonian. The Fourier
transformation of the vector potential is
\begin{equation}
A(p,t)= \frac{i A(t)}{p - i \eta}
\end{equation}
and $\eta$ is a positive infinitesimal. The matrix representation
of the Hamiltonian becomes
\begin{equation}
H_{nm}(t)= \frac{p^2_n}{2} \delta_{nm} + \sqrt{w_n w_m} \frac{i}{4
\pi} A(t) \frac{p_n+p_m+A(t)}{p_n-p_m + i \eta} .
\end{equation}
Taking the limit $\eta \rightarrow 0$, we obtain
\begin{equation}
i \partial_t c_n(t)=\eps_n(t) c_n(t) + \sum_{m \ne n} V_{nm}(t)
c_m(t)
\end{equation}
with diagonal kinetic energies
\begin{equation}
\eps_n(t)=\frac{1}{2} \left(p_n + \frac{1}{2} A(t) \right)^2 +
\frac{1}{8} A^2(t)
\end{equation}
and off-diagonal time-dependent couplings
\begin{equation}
V_{nm}(t)=\frac{i}{4 \pi} \sqrt{w_n w_m} A(t)
\frac{p_n+p_m+A(t)}{p_n-p_m}, \quad p_n \ne p_m
\end{equation}
We propagate numerically the equations of motion with the
Crank-Nicholson method for small equidistant time steps $\delta=6$
as
\begin{equation}
\left( {\bf I} + i \frac{\delta}{2} {\bf H}(t) \right) \cdot {\bf
c}(t-\delta) = \left( {\bf I} - i \frac{\delta}{2} {\bf H}(t)
\right) \cdot {\bf c}(t),
\end{equation}
subject to the final conditions $c_n(p_f)=\delta_{nf}/\sqrt{w_f}$
specified in the remote future $t \rightarrow \infty$. The
off-diagonal matrix elements of the momentum are evaluated using
these mesh points
\begin{equation}
p_{fi}(t) = e^{-i \eps_i t} \sum_{n=1}^N \sqrt{w_n} c_n^{\ast}(t)
p_n \phi_i(p_n)
\end{equation}
For the strip  $[p_f-0.5,p_f+0.5]$, convergence is reached when
the number of basis states $N \ge 549$, here $p_f=\sqrt{2
(E_f-W)}$ is the central photoelectron momentum.

\section*{Acknowledgements}
The author thanks N.~Minkov and A.~Georgieva for providing a slot
to give a talk on "Attosecond streaking of electrons from metal
surfaces" at the 33rd International Workshop on Nuclear Theory
IWNT33-2014.

\end{document}